\documentclass[prl,twocolumn,showpacs]{revtex4}
\usepackage{amsfonts,amsmath,mathrsfs,epsfig,amsbsy,bm,verbatim}

\newcommand{\ket}[1]{\left\vert {#1} \right\rangle}
\newcommand{\expect}[1]{\langle {#1} \rangle}

\newcommand{\ua}{\uparrow}
\newcommand{\da}{\downarrow}

\begin{document}

\title{Topologically protected surface Majorana arcs and bulk Weyl fermions in ferromagnetic superconductors}
\author{Jay D. Sau$^1$}

\author{Sumanta Tewari$^{2}$}
\affiliation{$^1$Department of Physics, Harvard University, Cambridge, MA 02138 \\
$^2$Department of Physics and Astronomy, Clemson University, Clemson, SC
29634}

\begin{abstract}
A number of ferromagnetic superconductors have been recently discovered which are believed to be in the so-called
  ``equal spin pairing" (ESP) state. In the ESP state the Cooper pairs condense forming order parameters
    $\Delta_{\uparrow\uparrow}, \Delta_{\downarrow\downarrow}$ which are decoupled in the spin-sector. We show that these
    three-dimensional systems
     should generically support topologically protected surface Majorana arcs and bulk Weyl fermions as gapless excitations.
 Similar protected low-energy exotic quasiparticles should also appear in
     the recently discovered non-centrosymmteric superconductors in the presence of a Zeeman field. The protected surface arcs
       can be probed by angle-resolved photoemission (ARPES) as well as fourier transform
scanning tunneling spectroscopty (FT-STS) experiments.
\end{abstract}

\maketitle

\paragraph{Introduction:}
Ferromagnetic superconductors (FS) display a remarkable coexistence of the mutually exclusive order parameters of ferromagnetism and superconductivity in
 the same band of electrons \cite{Aoki-Review,Mineev-review}. In the recently discovered Uranium-based Ising ferromagnetic superconductors
 UGe$_2$ \cite{Saxena}, URhGe \cite{Aoki}, and UCoGe \cite{Huy}, it is believed  that superconductivity appears in the spin-triplet $p$-wave
 equal-spin-pairing channel \cite{Vollhardt}. In this channel the superconducting order parameters $\Delta_{\uparrow\uparrow}$ and
 $\Delta_{\downarrow\downarrow}$ are decoupled in the spin space. In FS typically the ferromagnetic transition temperature $T^*$ far exceeds the
 superconducting transition temperature $T_c$, and the complex interplay of the spin-triplet superconductivity and ferromagnetism
leads to a fascinating phase diagram as a function of temperature, pressure, and magnetic field. In this paper we show that in 3D the superconductivity
 in these materials generically supports protected chiral Dirac or Weyl nodes in the bulk and
topologically robust open Majorana fermion arcs on the surface. Similar bulk and surface gapless quasiparticles, but without the Majorana character,
 have been recently predicted in some non-superconducting systems \cite{Vishwanath,Xu,Burkov}. They have also been predicted to occur in
superfluid He3-A \cite{tsutsumi, volovik1} and the recently discovered inversion-asymmetric superconductors known as non-centrosymmetric superconductors
 (NCS) \cite{Togano,Badica,Amano} in the presence of time reversal (TR) invariance \cite{Andreas}. As we show below, the FS systems are another candidate
 providing one of the simplest platforms
for the realization and detection of 3D gapless topological superconductivity (TS) with protected surface Majorana quasiparticles. Such
 modes on the surface should
be accessible to surface sensitive probes Fourier transform scanning tunneling spectroscopy (FT-STS)
 \cite{seamus,balatsky,yazdani}.

 The Weyl nodes in the FS are characterized by an energy dispersion linearly
proportional to momentum and come with a specific handedness or chirality. There are an even number of such nodes with each pair
 consisting of
nodes with opposite chiralities. Such chiral Dirac, or Weyl, nodes are topologically protected in 3D by an invariant \cite{Volovik}
 which takes the values
$\pm 1$, and hence can only be removed when a pair of such nodes with opposite signs of the invariant collide in the momentum space.
From the bulk-boundary correspondence such topologically protected gapless bulk spectrum leads to open Majorana fermion arcs on the surface which
 should be detectable in STM tunneling experiments.

Recently, there has been a lot of excitement about two- and one-dimensional TS systems with broken TR symmetry which can be used to
realize interesting non-Abelian quasiparticles such as Majorana fermions. Zero-energy Majorana fermions, defined by
hermitian operators $\gamma_0^{\dagger}=\gamma_0$, can in principle be used to realize fault-tolerant architecture for
 quantum computation \cite{RMP}. The canonical
 example of these systems is the
 2D spin-less $p_x+ip_y$ superconductor \cite{Read} which is gapped in the bulk and has a single zero-energy Majorana fermion mode localized
 at order parameter
  defects such as vortices. Additionally, this system has gapless chiral Majorana fermion modes
(defined by operators $\gamma_k^{\dagger}=\gamma_{-k}$) on the sample boundary. The robustness of such non-trivial
 topological properties of a 2D spin-less $p_x+ip_y$ superconductor can be understood in terms of the existence of
 a $Z_2$ topological invariant which can be viewed as the parity of the
Chern number of the corresponding 2D particle-hole symmetric  Bogoliubov-de Gennes (BdG) Hamiltonian \cite{Kitaev,parag}.
    Practical realizations of these exotic properties have recently been proposed in strong topological insulators (TI)
     which have a strong spin-orbit coupling \cite{Fu-Kane-Majorana} and, remarkably, even in ordinary semiconductors with
     a strong spin-orbit coupling and a suitably directed Zeeman splitting \cite{Sau-2010,Long-PRB,Oreg}.

    In this paper we extend the concepts of TR-breaking TS states
  to 3D solid state materials and investigate possible experimentally accessible condensed matter systems where such physics can be realized.
 We find that FS systems
 are ideal candidates in 3D where very similar physics is realized
\emph{even in the absence of a spin-orbit coupling and an external Zeeman field}. Interestingly, we find that in FS the analogous
 states are no-longer gapped in the bulk, but have gapless points in the momentum space  with a Weyl spectrum of the BdG quasiparticles.
Even more interestingly, the existence of the bulk Weyl nodes directly corresponds to the existence of open gapless Majorana fermion
 arcs on some suitable surfaces of the 3D system. The open surface Majorana fermion arcs offer the tantalizing possibility of detecting gapless Majorana excitations using the available surface sensitive probes such as ARPES and STM tunneling experiments.
\paragraph{Topological phases in ferromagnetic superconductors:}
Ferromagnetic superconductors, unlike conventional $s$-wave superconductors,
are characterized by a spin-triplet pairing potential which separates in the spin-sector
 (i.e., $\Delta_{\uparrow\uparrow}$ and $\Delta_{\downarrow\downarrow}$ are the relevant order parameters). Moreover,
because of the existence of a strong internal TR-breaking exchange field far greater than the usual Pauli limit, the superconducting
order is thought to be \cite{ohmi}  of the so-called non-unitary $p$-wave type \cite{Vollhardt}.  Based on these constraints, we will take below the relevant
 representative order parameter as $\Delta_{\sigma\sigma'}(\bm k)=\delta_{\sigma\sigma'}\Delta_{\sigma}\frac{(k_x+ik_y)}{k_F}$,
 with $\Delta_{\uparrow}\neq \Delta_{\downarrow}$, which is the order parameter of the $A_2$ phase of He3 \cite{Vollhardt}. The particular orbital
 form of the order parameter, however, will only be used at the end for numerical calculations. Most of the general results below
 (including the important predictions of the bulk Weyl modes and surface Majorana arcs) follow more generally simply from the
 equal-spin-pairing structure of the superconductivity of these materials. This is important because so far
 there is no consensus on the orbital symmetry of the superconducting order in FS.

 The mean-field Hamiltonian describing an ESP state for a $p$-wave superconductor is written as
\begin{align}
&H_{BCS}=\sum_{\sigma}\int d^2\bm k (\frac{k_x^2+k_y^2+k_z^2}{2 m^*}-\varepsilon_{F,\sigma})f^\dagger_{\sigma\bm k}f_{\sigma\bm k}\nonumber\\
&+[\Delta_{\sigma\sigma'}(\bm k)f^\dagger_{\sigma\bm k}f^\dagger_{\sigma'-\bm k}+h.c]\label{bcsferro}
\end{align}
where $\sigma=\ua,\da$ labels the spin index of the electron operators and the pairing
 potential is odd in momentum space so that $\Delta_{\sigma\sigma'}(\bm k)=-\Delta_{\sigma'\sigma}(-\bm k)$.
The magnetization of the ferromagnetic superconductor is accounted for by the difference in the
Fermi energies $(\varepsilon_{F,\ua}-\varepsilon_{F,\da})$.
Defining the Nambu spinor $\Psi(\bm k)=(\psi^\dagger_{\ua}(\bm k),\psi^\dagger_{\da}(\bm k),\psi_{\ua}(-\bm k),\psi_{\da}(-\bm k))$, the BdG Hamiltonian for the ferromagnetic SC is written as
\begin{align}
&H_b(k_x,k_y,k_z)=(\frac{k_x^2+k_y^2+k_z^2}{2 m^*}-\varepsilon_{F,av}-M N(0)\sigma_z)\tau_z\nonumber\\
&+[\Delta(\bm k)\tau_++h.c]\label{hesp}
\end{align}
where $M=\frac{\varepsilon_{F,\ua}-\varepsilon_{F,\da}}{2 N(0)}$ is proportional to the
magnetization, $N(0)$ is the density of states at the fermi level, and
 $\varepsilon_{F,av}=\frac{(\varepsilon_{F,\ua}+\varepsilon_{F,\da})}{2}$ is the average fermi energy of the two spin components.

The properties of the three-dimensional phase described by Eq.~\ref{hesp}  can be understood by dimensional
 reduction in momentum space to classes of two-dimensional topological phases. The idea of dimensional reduction
is to consider the Hamiltonian $H_b(k_x,k_y,k_z)$ for the translationally invariant 3D bulk system
 as a set of 2D systems i.e.,
\begin{equation}
H_{k_z}^{(2D)}(k_x,k_y)=H_b(k_x,k_y,k_z)\label{dimred}
\end{equation}
where $H_{k_z}^{(2D)}(k_x,k_y)$ are effective 2D Hamiltonians parametrized by $k_z$.
The topological properties of such 2D Hamiltonians, which break the TR symmetry,
are characterized by the Chern number associated with the occupied states \cite{chernnumber}.
The Chern number of a 2D topological Hamiltonian counts the number of chiral
edge states at a given $k_z$. The edge states of the 2D system can be understood
as surface states of the 3D system for surfaces which are parallel to the $z$-axis.
For such surface states, the wave-vector along $z$, $k_z$ is a good quantum
number and the dispersion of the surface states can be obtained from the dispersion
of the edge states.

\paragraph{Pfaffian topological invariants:}
The 3D BdG Hamiltonian $H_b$ of the FS systems has a particle-hole symmetry $\Lambda=i\tau_y K$, which anti-commutes with $H_b$. Here $K$ is the complex conjugation operator. To make use of this symmetry, we will assume from here onwards that $H_b$ is an
even function of $k_z$ so that $H_{k_z}^{(2D)}$ is also particle-hole symmetric.
In this case, one can use the Pfaffian topological invariant, $\textrm{sgn}(Pf(i\tau_y H_{k_z}^{(2D)}(k_x=k_y=0)))$
 \cite{parag} to determine the parity of the Chern number, which
must change whenever the Pfaffian changes sign. Moreover, since the pairing
potentials vanishes at $(k_x=k_y=0, k_z)$, the Pfaffian topological
invariant is found to be
\begin{equation}
\textrm{sgn}(Pf(i\tau_y H_{k_z}^{(2D)}(k_x,k_y=0)))=\textrm{sgn}(\varepsilon_{F,\ua}-k_z^2)\textrm{sgn}(\varepsilon_{F,\da}-k_z^2).
\end{equation}
 The above topological invariant is non-trivial only in the restricted range
of values of $k_z$ which satisfy:
\begin{equation}
k_{z,c,\da}=\sqrt{2 m^*\varepsilon_{F,\da}}<|k_z|<k_{z,c,\ua}=\sqrt{2 m^*\varepsilon_{F,\ua}}.\label{toprangeferro}
\end{equation}
For values of $k_z$ outside this range, the system has even Chern parity.

\paragraph{Bulk Weyl fermions:}
The family of topological superconductors described by the Hamiltonian $H_{k_z}^{(2D)}$
undergoes a quantum phase transition from the topological to the non-topological phase when the Chern parity changes
at the values of $k_z$ where the conditions in Eq.~\ref{toprangeferro} are saturated. Such topological quantum
phase transitions are accompanied by a closing of the topological gap at $k_x=k_y=0$.
The set of energy eigenvalues of the 3D Hamiltonian $H_b(k_x,k_y,k_z)$ in Eq.~\ref{hesp} is a union of the
energy eigenvalues of the entire family of 2D topological superconductors described by the Hamiltonians $H_{k_z}^{(2D)}$.
 Therefore the Hamiltonian $H_b(k_x,k_y,k_z)$ must
have gapless points at $\bm K=(k_x=0,k_y=0,k_z=\pm k_{z,c,\sigma})$, with a two-fold degeneracy of eigenstates, $\ket{\tau_z=\pm 1,\sigma_z=\sigma}$,
 where $\sigma=\ua,\da$.
The dispersion of the pair of degenerate states ( $\ket{\tau_z=\pm 1,\sigma_z=\sigma}$)
 around the degeneracy points $\bm K$ can be obtained by
expanding the Hamiltonian in Eq.~\ref{hesp} as $\bm k=\bm K+\delta\bm k$
 to linear order in $\delta\bm k$.
The resulting Hamiltonian takes the form of a three-dimensional Dirac cone,
\begin{align}
&H_{b}(\bm K+\delta\bm k)=\bm {\delta k}\cdot \bm \nabla \Delta_R(\bm K) \tau_x+\bm {\delta k}\cdot \bm \nabla \Delta_I(\bm K) \tau_y\nonumber\\
&+\delta k_z \frac{2 k_{z,c,\sigma}}{m^*}\tau_z+o(\delta \bm k^2),\label{dirac}
\end{align}
where $\Delta_{R,I}(\bm K)$ are the real and imaginary parts of $\Delta(\bm K)$ in Eq.~\ref{hesp}.
Such Dirac cones in 3D are  protected because continuous deformations of the
bulk Hamiltonian $H_{b}(\bm k)$, which can only redefine $\bm {\delta k}$ in Eq.~\ref{dirac},
 cannot gap out such a Dirac cone.
A perturbation to Eq.~\ref{hesp} can only get rid of Dirac cones from the spectrum by shifting and merging
 them in pairs \cite{Volovik}. This kind of protection of Dirac cones in $D=3$ 
 can be further represented by associating it with a topological
invariant \cite{Volovik}. Bulk semi-metals in three dimensions with such Dirac-like point fermi surfaces are
referred to as Weyl semimetals.
Thus $H_b$ in Eq.~\ref{hesp} represents a nodal superconductor which is
the analog of a Weyl semimetal with four isolated Dirac cones.

\paragraph{Surface Majorana arcs:}
The 2D topological superconductor Hamiltonians $H_{k_z}^{(2D)}(k_x,k_y)$, which are parameterized by $k_z$, have an odd
Chern number in the range Eq.~\ref{toprangeferro} and are therefore characterized by
chiral Majorana edge states that are confined to the edge of the system. A surface along the $x-z$ plane
for the original 3D Hamiltonian Eq.~\ref{hesp} is translationally invariant along the $z$ and $x$ directions and therefore has
well-defined $k_z$ and $k_x$ momenta. Therefore the surface state with a fixed $k_z$ and $k_x$ of the 3D Hamiltonian Eq.~\ref{hesp}
is identical to the edge state with momentum $k_x$ of the 2D Hamiltonian in Eq.~\ref{dimred} with  $k_z$ as
a parameter value. The energy of such a chiral Majorana mode vanishes for $k_x=0$ at
any value of $k_z$ in the range in Eq.~\ref{toprangeferro} and therefore appears on the surface ARPES spectrum as a Majorana
arc.

The typical extent in $k_z$ of such a non-degenerate Majorana arc is limited to the range given in Eq.~\ref{toprangeferro}.
It is important to note that in contrast to TR breaking TS states in $D=2,1$, where edge  Majorana fermions appear only in a
 restricted range of parameters (for example, for chemical potential smaller than the Zeeman coupling in Refs.~\cite{Sau-2010,Long-PRB,Oreg}),
 in 3D FS systems Majorana arcs appear \emph{for any value of the chemical potential}
 (albeit in a restricted range in momentum space, Eq.~\ref{toprangeferro}).
Below, as mentioned earlier, we take the order parameter of the FS systems to be of the non-unitary ESP type with
 an orbital structure given by
$\Delta_{\sigma\sigma'}(\bm k)=\delta_{\sigma\sigma'}\Delta_{\sigma}\frac{(k_x+i k_y)}{k_F}$ with $\Delta_{\ua}\neq\Delta_{\da}$.
In this case, the pairing potential
$\Delta_{\sigma\sigma'}(\bm k)$ is associated with a Chern number 2,  therefore one would have a pair of chiral surface modes
 propagating along the surface in the range $|k_z|<k_{z,\da}$.
This could in general appear as a pair of fermi arcs of Bogolibov quasiparticles in the ARPES or STM spectrum.
Since the BdG Hamiltonian now decouples into a spin-$\ua$ and spin-$\da$ sector, the 2D Hamiltonian $H_{k_z,\sigma}^{(2D)}(k_x,k_y)$
for each spin $\sigma$ can be thought of as independent odd Chern number topological superconductors which has a
Majorana chiral edge mode in the ranges
\begin{equation}
|k_z|<k_{z,c,\ua}\label{toprangeferroup}
\end{equation}
for spin-up electrons and
\begin{equation}
|k_z|<k_{z,c,\da}\label{toprangeferrodn}
\end{equation}
for spin-down electrons.
Therefore we find that the surface arcs exist over a much larger range in $k_z$ i.e between $-k_{z,c,\ua}$ and $k_{z,c,\ua}$.

These chiral edge modes exist in the spin-$\sigma$ sector for $k_z$ satisfying $|k_{z,\sigma}|<k_{z,c,\sigma}$ and have a dispersion
of the form $\varepsilon_{k_z}(k_x)=v(k_z)k_x$. Therefore the dispersion of the corresponding surface mode is given by
\begin{equation}
\epsilon(k_x,k_z,\sigma)=v(k_z,\sigma)k_x\sim \frac{\Delta_\sigma(k_z)}{\sqrt{2 m^* (\varepsilon_{F,\sigma}-k_z^2)}}k_x\label{surfacedisp}
\end{equation}
for $|k_{z}|<k_{z,c,\sigma}$
and
\begin{equation}
|\epsilon(k_x,k_z,\sigma)|>\epsilon_g
\end{equation}
where $\epsilon_g$ is a finite positive gap for $|k_{z}|>k_{z,c,\sigma}$.
The dispersion of the surface modes given in Eq.~\ref{surfacedisp} has the special property that
the energy vanishes on a pair of lines $k_x=0$ which terminates at $k_{z}=\pm k_{z,c,\sigma}$. These
lines are referred to as Majorana arcs.

As mentioned before the Majorana character of the surface arcs is only protected in the restricted range in Eq.~\ref{toprangeferro}.
In the rest of the $k_z$ range, the dispersion arcs are not strictly speaking Majorana in character. In particular an in-plane
Zeeman splitting along $x$, $V_{Z,x}$, can lead to a mixing of the $\sigma=\ua,\da$ states so that the surface Majorana arcs now split into a
pair of fermi arcs, indexed by $s=\pm 1$, with dispersion
\begin{align}
&\epsilon(k_x,k_z,s)\sim \frac{v(k_z,\ua)+v(k_z,\da)}{2}k_x\nonumber\\
&+s\sqrt{\left(\frac{v(k_z,\ua)-v(k_z,\da)}{2}\right)^2 k_x^2+V_{Z,x}^2}\label{surfacedisp1}
\end{align}
for $|k_z|<k_{z,c,\da}$.
However, one must note that while these modes are not Majorana in the strict sense, they still have equal weights of electron
and hole states.

The Majorana arcs are obtained by solving $\epsilon(k_x,k_z,s)=0$. This leads to the equation for the Majorana arc
\begin{equation}
k_x=\pm\frac{V_{Z,x}}{\sqrt{v(k_z,\ua)v(k_z,\da)}}\approx \pm\frac{ V_{Z,x} \sqrt{\epsilon_F-k_z^2}}{\sqrt{\Delta_{\ua}\Delta_{\da}}}\label{eq:fermiarcs}
\end{equation}
where we have assumed the magnetization to be small compared to the total density (i.e. $|\varepsilon_{\ua}-\varepsilon_{\da}|\ll \varepsilon_{F,av}$) and the expression $\Delta_{\sigma}(k_z)=\Delta_{\sigma}$. The resulting Majorana arcs are ellipses in the $(k_x,k_z)$ plane
between $|k_z|<k_{z,c,\da}$  as plotted in Fig. 1.

\begin{figure}
\centering
\includegraphics[scale=0.3,angle=0]{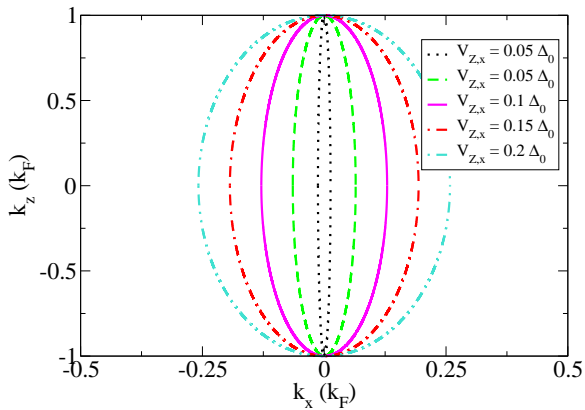}
\caption{ $V_{Z,x}$-dependent surface fermi contour plotted according to
 Eq.~\ref{eq:fermiarcs} for $\Delta_{\da}=0.6\Delta_{\ua},\Delta_{\ua}=\Delta_0$. The width of the contour in the
$k_x$ direction is proportional to $V_{Z,x}$. As $V_{Z,x}\rightarrow 0$, the spin-$\ua$ and spin-$\da$ sectors
decouple and the fermi contour evolves into a pair of surface Majorana arcs.
}\label{Fig1}
\end{figure}

\paragraph{Quasiparticle interference of Majorana arcs in STM:}
The surface Majorana arcs shown in Fig.~1
 should in principle be directly visible as arcs in momentum space
 at the Fermi energy in the ARPES spectrum.
However, since these arcs are separated from the bulk states by a relatively small superconducting gap ($\sim 1K\sim 0.1$ meV),
it is not clear if the typical energy resolution achieved in ARPES is sufficient to resolve the Majorana arcs. On the other hand,
low temperature and low noise STM measurements often have sub-100 mK energy resolution allowing one access to the
energy scale of the Majorana arcs. While conventional STM, which measures the local quasiparticle density of states at the surface,
would be able to detect the presence of Majorana arcs as gapless modes at the surface, it does not provide any information about
the structure (finite extent in momentum, curvature, etc.) of the Majorana arcs in momentum space.
 Such information can be obtained by FT-STS
 \cite{seamus,balatsky,yazdani}.
FT-STS relies on the fact that impurity-scattering at the surface leads to a spatially
varying local quasiparticle density of states, $n(\bm r)$ at the surface, which can be determined from the
spatial variation of the tunneling current $I_{t}(\bm r)\propto n(\bm r)$ at the surface. The resulting current map
obtained from STM can be Fourier transformed to obtain $I_t(\bm q)=\int d\bm r I_t(\bm r)e^{i\bm q\cdot\bm r}\propto \int d\bm r n(\bm r)e^{i\bm q\cdot\bm r}$. The disorder averaged square of the Fourier transform $\expect{|I(\bm q)|^2}$ can be shown to be related to the
joint density of states $\rho(\bm q)$ (i.e. $\rho(\bm q)\propto\expect{|I(\bm q)|^2}$) defined by
\begin{equation}
\rho(\bm q)=\int d\bm k \delta(\epsilon(\bm k))\delta(\epsilon(\bm k+\bm q))\label{eq:rho},
\end{equation}
where $\epsilon(\bm k)$ is the surface mode dispersion in Eq. 11.
\begin{figure}
\centering
\includegraphics[scale=0.4,angle=0]{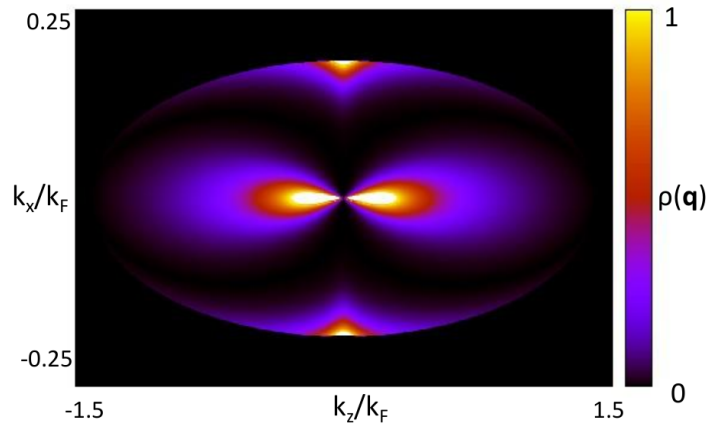}
\caption{(color online) The joint density of states $\rho(\bm q)$ (in arbitrary units) at the fermi surface,
 which, as explained in the text, would be measured by FT-STS on the surfaces of FS systems.
 We have used Eq. ~\ref{eq:rho}, and the parameter values $V_{Z,x}=0.1\Delta_{\ua}$ and $\Delta_{\da}=0.6\Delta_{\da}$
 (i.e. same as Fig.~\ref{Fig1}).
}\label{Fig2}
\end{figure}
The application of the FT-STS method \cite{seamus,balatsky,yazdani} outlined above for the weakly disordered surface
of a ferromagnetic semiconductor is expected to lead to the characteristic structure
 that is plotted in Fig.~\ref{Fig2}.

\paragraph{Summary and Conclusion:}
We propose the recently discovered ferromagnetic superconductors \cite{Aoki-Review,Mineev-review,Saxena,Aoki,Huy} as
experimentally accessible 3D systems where the physics of TR-breaking TS states is realized. These systems support
robust bulk gapless quasiparticles in the form of chiral Weyl fermions and protected open Majorana fermion arcs on suitably
 oriented surfaces of the 3D system. In contrast to the TR-breaking TS states in $D=2,1$ \cite{Sau-2010,Long-PRB,Oreg} where Majorana modes
 appear only in a restricted range of the chemical potential smaller than an applied Zeeman field, remarkably, in the 3D systems
 considered here
 edge Majorana modes appear \emph{for all values of the chemical potential}. 
The Weyl nodes in the bulk are topologically protected because they separate momentum space regions where the
 BdG Hamiltonians have different values of a relevant topological invariant, the analog of the 2D Chern number. Therefore
 the Weyl nodes arise from topologically unavoidable closing of the quasiparticle gap at isolated points in the momentum space.
The existence of the bulk Weyl nodes directly corresponds to the existence of open gapless
 Majorana fermion arcs (Fig.~1) on suitable surfaces of the 3D system. The surface Majorana fermion arcs offer the
 tantalizing possibility of detecting gapless Majorana excitations using the available surface sensitive probes such as
 ARPES and spectroscopic STM experiments (Fig.~2).

 The FS systems are not the only materials which can support 3D gapless TS states with broken TR symmetry. The newly
 discovered 3D  NCS materials \cite{Togano,Badica,Amano} can also support such states in the presence of a sufficiently
 large Zeeman splitting. In NCS (e.g., Li$_2$Pd$_x$Pt$_{3-x}$B, Y$_2$C$_3$), because of an intrinsic spin-orbit coupling,
 the superconducting order parameter has an admixture of $s$-wave and $p$-wave components. In the presence of a Zeeman
 splitting larger than the $s$-wave component of the order parameter, the NCS enters into a TR-breaking TS state with bulk Weyl
 fermions and surface Majorana arcs. The Zeeman splitting, however, should not be accompanied by a large orbital depairing field which may destroy the superconductivity itself. This can be ensured by choosing materials with a large enough $g$-factor so
 that a relatively small magnetic field can still create a Zeeman splitting larger than the $s$-wave part of the order parameter.

J.S. thanks the Harvard Quantum Optics Center for support. S.T. thanks DARPA and NSF for support.


\end{document}